\documentclass[amssymb,nobibnotes,aps,pre,reprint,superscriptaddress]{revtex4-2}
\usepackage[utf8]{inputenc}

\usepackage{natbib}
\usepackage{graphicx}
\usepackage{amsmath}
\usepackage{dcolumn} 
\usepackage[ruled, longend]{algorithm2e}
\SetAlgoCaptionLayout{raggedright} 

\newcommand{\N}{\mathbb N}

\newcommand{\R}{\mathbb R}
\newcommand{\cI}{\mathcal I}

\newcommand{\cL}{\mathcal L}
\newcommand{\eps}{\varepsilon}
\renewcommand{\d}{\partial}

\newcommand\argleft{\mathopen{}\mathclose\bgroup\left}
\newcommand\argright{\aftergroup\egroup\right}

\newcommand*\diff{\mathop{}\!\mathrm{d}}
\DeclareMathOperator{\sign}{sgn}

\graphicspath{{./}}

\begin{document}

\preprint{APS/123-QED}

\title{Sampling Rare Event Energy Landscapes via Birth-Death Augmented Dynamics}

\author{Benjamin Pampel}
\affiliation{Max Planck Institute for Polymer Research, Ackermannweg 10, 55128 Mainz, Germany}

\author{Simon Holbach}
\email{s.holbach@uni-mainz.de}
\affiliation{Institut f\"{u}r Mathematik, Johannes Gutenberg-Universit\"{a}t Mainz, Staudingerweg 9, 55099 Mainz, Germany}

\author{Lisa Hartung}
\email{lhartung@uni-mainz.de}
\affiliation{Institut f\"{u}r Mathematik, Johannes Gutenberg-Universit\"{a}t Mainz, Staudingerweg 9, 55099 Mainz, Germany}

\author{Omar Valsson}
\email{omar.valsson@unt.edu}
\affiliation{Max Planck Institute for Polymer Research, Ackermannweg 10, 55128 Mainz, Germany}
\affiliation{Department of Chemistry, University of North Texas, Denton, TX, USA}

\begin{abstract}
A common problem that affects simulations of complex systems within the computational physics and chemistry communities is the so-called sampling problem or rare event problem where proper sampling of energy landscapes is impeded by the presences of high kinetic barriers that hinder transitions between metastable states on typical simulation time scales.
Many enhanced sampling methods have been developed to address this sampling problem and more efficiently sample rare event systems.
An interesting idea, coming from the field of statistics, was introduced in a recent work (Y. Lu, J. Lu, and J. Nolen, arXiv:1905.09863, 2019) in the form of a novel sampling algorithm that augments overdamped Langevin dynamics with a birth-death process.
In this work, we expand on this idea and show that this birth-death sampling scheme can efficiently sample prototypical rare event energy landscapes, and that the speed of equilibration is independent of the barrier height.
We amend a crucial shortcoming of the original algorithm that leads to incorrect sampling of barrier regions by introducing a new approximation of the birth-death term.
We establish important theoretical properties of the modified algorithm and prove mathematically that the relevant convergence results still hold.
We investigate via numerical simulations the effect of various parameters, and we investigate ways to reduce the computational effort of the sampling scheme.
We show that the birth-death mechanism can be used to accelerate sampling in the more general case of underdamped Langevin dynamics that is more commonly used in simulating physical systems.
Our results show that this birth-death scheme is a promising method for sampling rare event energy landscapes.
\end{abstract}

\maketitle

\section{Introduction}

A common task in statistics, Bayesian inference, and machine learning is to sample a probability distribution~$\pi$ using sampling algorithms such as Monte Carlo~\cite{metropolis_equation_1953,hastings_monte_1970}, or Langevin dynamics~(LD)~\cite{langevin_theorie_1908,turq_brownian_1977}. However, it can be challenging to efficiently sample the probability distribution if it is multi-modal and exhibits metastability. Then, the transition time to go between different high-probability modes is long compared to the simulation times that one can employ. In other words, a transition between modes is a rare event. This sampling problem has led to the development of a wide range of advanced sampling algorithms to more efficiently sample probability distributions~\cite{sherman_monte_1986,liu_stein_2016,lu_accelerating_2019,10.1137/21m1425062}

A similar sampling or rare event problem is well known in the computational physics and chemistry communities~\cite{10.1039/d1cp04809k,valsson_enhancing_2016,Henin_Lelievre_Shirts_Valsson_Delemotte_2022}. There, one is interested in using atomistic molecular dynamics~\cite{alder_studies_1959} simulations to understand the behavior of physical systems. For example, this could be the formation and growth of a crystal~\cite{10.1021/acs.chemrev.5b00744}, the folding of a protein~\cite{10.1016/j.sbi.2011.12.001}, unbinding of a ligand from a protein complex~\cite{10.3389/fmolb.2022.899805}, and so forth. The rare event problem is viewed in terms of an energy landscape, given by the negative logarithm of the corresponding probability distribution, which is characterized by metastable states separated by high kinetic barriers that hinder transitions between states on typical simulation time scales. This energy landscape is called a free energy landscape if one considers the low-dimensional description of the system and its dynamics in the terms of so-called collective variables that capture the slow modes of the physical process. This rare event sampling problem has lead to the development of a wide range of so-called enhanced sampling methods within the molecular simulation field~\cite{
valsson_variational_2014,
valsson_enhancing_2016,
10.1039/d1cp04809k,
Henin_Lelievre_Shirts_Valsson_Delemotte_2022,
torrie_nonphysical_1977,
huber_local_1994,
Darve-JCP-2001,
Hansmann-PRL-2002,
Kastner2011umbreallsampling,
Maragakis-JPCB-2009,
laio_escaping_2002,
barducci_welltempered_2008,
Whitmer_BFS_2014,
Invernizzi2020opus,
giberti2021atlas}

A common sampling strategy is to consider multiple independent simulations that are started from different initial conditions. Each simulation then explores a different area of the energy landscape and by pooling the simulations together, one can obtain improved sampling statistics and results~\cite{coveney_calculation_2016,grossfield_best_2019}. However, for rare event systems, each independent simulation will still suffer from the same sampling issues due to a lack of transitions between metastable states, which will skew the sampling statistics and lead to incorrect results when the simulations are pooled together.

We can consider each independent simulation as a walker or a particle exploring an energy landscape. Thus, we can then view multiple independent simulations as an ensemble of independent particles that explore an energy landscape. To overcome the sampling problem, we can introduce some kind of interaction between the particles, for example by considering population dynamics for the particles, as has been done in various ways in different fields~\cite{anderson_random_1975,sherman_monte_1986,gilks_adaptive_1994,aldous_go_1994,10.1016/s0006-3495(96)79552-8,grassberger_go_2002,10.1063/1.3456985,10.1021/cr2001564,10.1021/cr2001564,10.1146/annurev-biophys-070816-033834,liu_stein_2016,rotskoff_global_2019,lu_accelerating_2019,10.1137/21m1425062,10.1063/1.2711185,10.1021/ct900524t}. Multiple walkers are also routinely combined with other enhanced sampling methods to accelerate convergence~\cite{raiteri_efficient_2006,10.1063/1.2711185,10.1021/ct900524t,valsson_welltempered_2015}.

In Ref.~\citenum{lu_accelerating_2019}, the authors introduce an interesting algorithm for sampling multi-modal probability distributions that augments overdamped Langevin dynamics with a birth-death process. Theoretically, this sampling scheme is formulated in terms of a Fokker-Planck-Birth-Death equation that adds a birth-death term to the conventional Fokker-Planck equation. In practice, the scheme is formulated in terms of a set of particles, each diffusing on an energy landscape according to overdamped Langevin dynamics, but also interacting with each other via non-local moves that are determined by an approximation of the birth-death term. It is shown that this scheme greatly improves the sampling and leads to considerably faster convergence to the equilibrium probability distribution in comparison to overdamped Langevin dynamics without a birth-death process.

In this work, we explore the potential of this birth-death sampling scheme and show that it can efficiently sample prototypical rare event energy landscapes. We show that the original algorithm in Ref.~\citenum{lu_accelerating_2019} suffers from a deficiency that leads to incorrect sampling of barrier regions, and we amend this shortcoming by introducing an alternative approximation of the birth-death term.
We establish important theoretical properties of the associated interacting particle systems and prove mathematically that the relevant convergence results still hold with this new approximate birth-death term. Furthermore, we show that the birth-death mechanism can be used to accelerate sampling in the general case of underdamped Langevin dynamics that is more commonly used in simulating physical systems.

In Section~\ref{sec:overview}, we introduce the fundamental idea behind the method. Section~\ref{sec:theory} provides the theory behind the birth-death scheme and our new approximation of the birth-death term along with mathematical proofs. In Section~\ref{sec:implementation}, we present the algorithm and details on the implementation. In Section~\ref{sec:applications}, we show applications to prototypical rare event energy landscapes and investigate the effect of various parameters of the algorithm. Finally, in Section~\ref{sec:summary}, we end with a few concluding remarks.

\section{Overview of the method}\label{sec:overview}

Before giving a formal theoretical description of the birth-death sampling scheme in the following section, we will provide here a simplified discussion of the basic mechanism of the method.

\subsection{Motivation and context}

The general starting point is a high-dimensional dynamical system whose dynamics we can describe precisely by propagating the system in time using molecular dynamics or Langevin dynamics simulations (or even Monte Carlo simulations). In order to better understand its metastable states and the transitions between them, it can however be much more insightful to study them in terms of only a few degrees of freedom of the system that capture the essential features one is interested in (so-called collective variables, see e.g., \cite{fiorin_using_2013,pietrucci_strategies_2017,10.1016/j.sbi.2017.02.006,valsson_enhancing_2016,Henin_Lelievre_Shirts_Valsson_Delemotte_2022}). In an atomistic simulation, for example, this might be the distances between selected atoms, or some dihedral angles, but also possibly more sophisticated quantities. Unfortunately, one typically cannot calculate the low-dimensional dynamics analytically from the full high-dimensional dynamics. However, it is possible to simulate the full system, track the low-dimensional degrees of freedom that are of interest, and then estimate features of the low-dimensional dynamics on this basis.

In this work, we will assume that the low-dimensional dynamics can be described by (overdamped) Langevin dynamics with respect to an energy landscape $U$. This is justified in practice if there is timescale separation between the slow degrees of freedom that define the energy landscape and the system's other degrees of freedom, in other words, they are adiabatically separated~\cite{Zwanzig_2001}.

We will also make the (artificial) assumption that $U$ is known a priori and therefore perform all simulations directly with respect to the low-dimensional space. This is typically not the case in practice (as there is no need to estimate $U$ when it is known), and our work should be viewed as a first step in which we check how accurately and how fast our proposed sampling algorithm can estimate prototypical reference energy landscapes. Adjusting our algorithm in such a way that it can handle real applications will be the content of future work.

\subsection{Description of the method}

We address the sampling problem in a setup where we have multiple independent simulations, each of which we interpret as the trajectory of a particle. We consider an ensemble of $N$ particles where each of them diffuses independently on the energy landscape $U$. Corresponding to $U$ is an equilibrium distribution $\pi$ that denotes the particle density in equilibrium. The energy landscape and the equilibrium distribution are related via the Boltzmann factor (see Eq.~\eqref{eq:pidef} below), so knowing (or estimating) $U$ is equivalent to knowing (or sampling) $\pi$. In order to sample $\pi$ (and, equivalently, estimate $U$) from the simulated data, we can consider two different averages as well as their combination: the time-average and the particle average.

For time-averaging, we consider an accumulated histogram, i.e., for a given particle we count how often it is observed at each position. In the limit of infinitely long simulation times this converges to the equilibrium distribution $\pi$. However, in rare event systems where energy barriers between metastable states are so high that transitions between them rarely occur on simulation time scales, a single particle will likely only explore one of the metastable states (depending on its initial position). Alternatively, we can for a given point in time average over the particles, i.e., we look at the current particle distribution in order to approximate $\pi$. By considering the whole ensemble of independent simulations, and combining the two averages into the ensemble average (i.e., averaging over the particles and time), we would hope to obtain better sampling statistics and results, but a lack of transitions between metastable states will typically still result in poor estimates of the barrier regions and the energy differences between different metastable states.

The birth-death sampling scheme that we present in this work aims to improve this situation by introducing non-local particle moves. The basic idea is that, if there is a significant difference between the particle distribution and the equilibrium one at some given time, particles are killed and duplicated in such a way that this spatial redistribution reduces the deviation instantaneously.

\begin{figure*}[htb]
    \includegraphics[width=\textwidth]{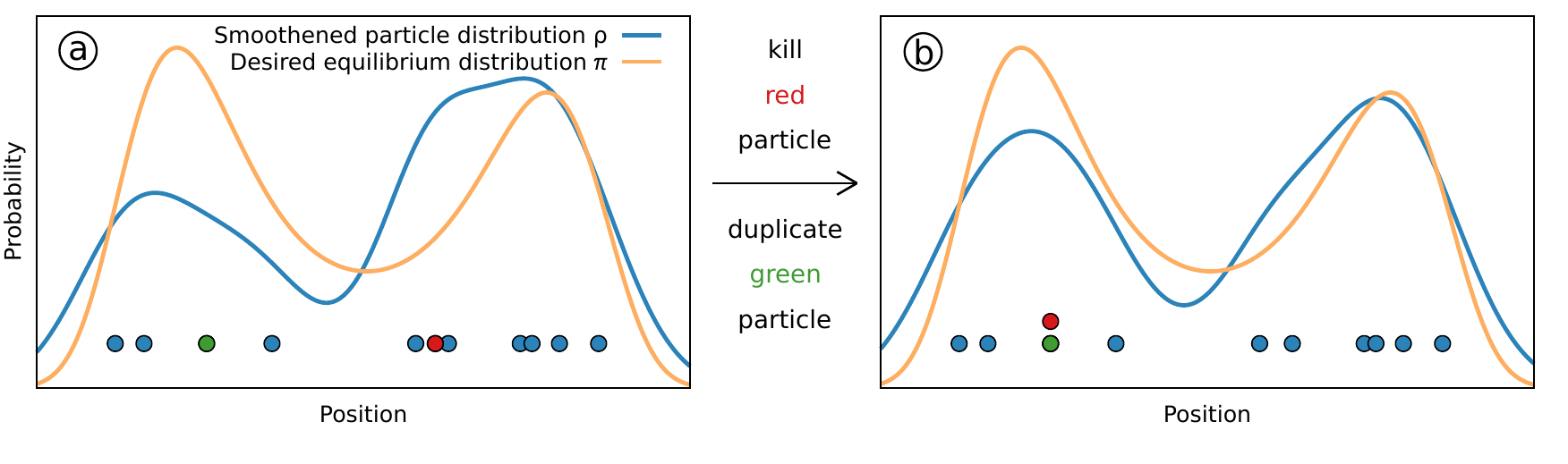}
    \caption{Sketch of the method: (a) positions of particles (height irrelevant) and a smooth approximation of their ensemble particle distribution $\rho(x)$ (blue line) together with the desired equilibrium distribution $\pi(x)$ (orange line). At the position of the green particle the current ensemble particle distribution is significantly lower than the desired equilibrium distribution, while it is higher at the position of the red particle.
    (b) New ensemble particle distribution after the red particle of (a) has been killed and the green particle been duplicated. Effectively, this means that the red particle has performed a non-local move to the position of the green particle.
    \label{fig:sketch}}
\end{figure*}

To understand this idea a little better, let us consider the one-dimensional example shown in Fig.~\ref{fig:sketch}. Here, the energy landscape $U$ is a double-well potential, so the equilibrium distribution $\pi$ (orange line) features two local maxima with a local minimum in-between. To be able to compare the discrete distribution of the particles to the continuous equilibrium density $\pi$ more easily, we first smoothen it by placing centered Gaussian kernels at the particles' positions (similar to a Kernel density estimation~\cite{silverman_density_1998}) to arrive at the smoothened ensemble particle density $\rho$ (blue line). In panel (a), we can see a significant difference between the equilibrium distribution that we want to sample and the current ensemble particle distribution: the region to the left is undersampled (i.e., $\rho < \pi$) and the region to the right is oversampled (i.e., $\rho > \pi$).

To bring the two distributions in better agreement, we now propose to kill and duplicate particles. While in the actual algorithm this decision would be made in a stochastic manner, we here consider an exemplary move based on intuition. We want to move some mass of the particle density from the oversampled area on the right to the undersampled region on the left. To achieve this, we decide to kill the red particle and duplicate the green particle. By doing this, the red particle has basically performed a non-local move from its original position to the position of the green particle. Panel (b) of Fig.~\ref{fig:sketch} shows the resulting ensemble particle distribution where we can see that the two distributions now are in much better agreement. We have therefore managed to obtain a momentary particle distribution that is closer to the one in equilibrium. Thus, by performing such non-local birth-death moves from time to time while in-between the particles diffuse independently, we should quickly obtain an ensemble particle distribution that is in good agreement with the equilibrium distribution.

The crucial problem is now to perform these birth-death moves in such a way that we preserve global sampling statistics. Clearly, our manual killing and duplication of particles does not achieve this. We will therefore use a stochastic approach, where birth-death events occur at random times that depend on a slightly more involved comparison of the equilibrium distribution and the current local particle density (compare Eq.~\eqref{eq:birth-death-functional2} below). Analytically, this combination of Langevin dynamics and birth-death events can be expressed by a Fokker-Planck equation with an additional birth-death term, as will be introduced in the following section.

Recall that in practice $\pi$ is typically not known a priori, making it more difficult to determine birth-death events based on how much the current particle distribution differs from it. Adjusting the birth-death mechanism in such a way that it still works with an on-the-fly estimation of $\pi$ via enhanced sampling methods will be the content of future research.

\section{Theory}\label{sec:theory}

\subsection{Overdamped Langevin dynamics with a birth-death process}

The inertia-free motion of a particle with initial position $x(0)\in\R^d$ in the smooth potential (i.e., energy landscape) $U\colon\R^d\to\R$ is described by the overdamped Langevin equation
\begin{equation}\label{eq:SDE}
  \diff x(t) = -D \beta \nabla U(x(t))\diff t + \sqrt{2D} \diff W(t)
\end{equation}
where $D>0$ is the diffusion coefficient, $\beta= 1/k_{\mathrm{B}}T>0$ is the inverse thermal energy at temperature $T$ with the Boltzmann constant $k_{\mathrm{B}}$, and $W$ is a standard Brownian motion on $\R^d$. Its solution $X=(x(t))_{t\ge0}$ is a Markov process that has a unique stationary distribution, the density of which is given by
\begin{equation}\label{eq:pidef}
  \pi(x)=Z^{-1} \cdot \mathrm{e}^{-\beta U(x)},
\end{equation}
where $Z$ is a normalizing constant (generally called the partition function in statistical physics). A constant shift of the potential does not change the dynamics described by Eq.~\eqref{eq:SDE}. Furthermore, exact knowledge of the normalizing constant is not required for practical applications of the methodology described in this paper. Therefore, we may assume without loss of generality that $Z=1$. Then
\begin{equation}\label{eq:Uintermsofpi}
  U(x)=-\beta^{-1}\log\pi(x)
\end{equation}
and we will write everything in terms of $\pi(x)$ in the sequel, omitting the dependence on $\beta$. For functions of the space variable $x\in\R^d$, we will usually omit the argument unless it is specifically needed.

We would like to stress that by Eqs.~\eqref{eq:pidef} and \eqref{eq:Uintermsofpi}, knowing the energy landscape and knowing the equilibrium distribution is the same thing.

The differential operator $L$ that is the generator of $X$ and its formal adjoint $L^*$ are given by
\begin{equation}
	L f	=D \Delta f + D\nabla \log\pi \cdot \nabla f
\end{equation}
and
\begin{equation}
	L^* f	=D\Delta f - D\nabla\cdot\left(f \nabla \log\pi\right) =D\nabla\cdot\left(f \nabla \log\frac{f}{\pi}\right)
\end{equation}
for any smooth function $f\colon\R^d\to\R$. The transition densities $\rho_t\colon\R^d\to[0,\infty)$ of the process $X$ satisfy the corresponding (linear) Fokker-Planck equation
\begin{equation}\label{eq:Fokker-Planck}
	\d_t\rho_t=L^* \rho_t,
\end{equation}
to which $\pi$ is a stationary solution, i.e.,
\begin{equation}\label{eq:stationary-solution}
	L^*\pi=0.
\end{equation}
If $f\colon\R^d\to\R$ is a probability density function and $g\colon\R^d\to\R$ is a formal probability density function (i.e., a convex combination of a probability density function in the classical sense and delta-distributions), we define the unitless function $\alpha_\pi (f,g)\colon\R^d\to\R$ with
\begin{equation}\label{eq:birth-death-functional}
	\alpha_\pi (f,g)(x)=\log\frac{f(x)}{\pi(x)}-\int\log\left(\frac{f(y)}{\pi(y)}\right)g(y) \diff y.
\end{equation}
We then define the (non-linear and non-local) Fokker-Planck-Birth-Death equation as
\begin{equation}\label{eq:Fokker-Planck-Birth-Death}
	\d_t\rho_t=L^* \rho_t -\tau_{\alpha} \alpha_\pi (\rho_t) \rho_t,
\end{equation}
where
\begin{equation}\label{eq:birth-death-functional2}
	\alpha_\pi (\rho_t)=\alpha_\pi (\rho_t,\rho_t)
\end{equation}
is called the birth-death term and $\tau_{\alpha}>0$ is a rate factor that has units of 1/time. Note that the rate factor was not included in the Fokker-Planck-Birth-Death equation in Ref.~\citenum{lu_accelerating_2019}, in other words, the authors assumed that $\tau_{\alpha}=1$.
We have that
\begin{equation}\label{eq:stationary-solution-birth-death}
	\alpha_\pi(\pi)=0
\end{equation}
and hence $\pi$ is also a stationary solution to the Fokker-Planck-Birth-Death equation (Eq.~\eqref{eq:Fokker-Planck-Birth-Death}). In other words, adding the birth-death term to the Fokker-Planck equation in this way does not change the equilibrium.

Note that the normalization constant of $\pi(x)$ cancels out in Eq.~\eqref{eq:birth-death-functional}, so one can apply the birth-death term without knowing the normalization of the stationary distribution.

\subsection{An interacting particle approach}\label{sec:theory_algorithm}

The overdamped Langevin equation (Eq.~\eqref{eq:SDE}) can be thought of as the probabilistic counterpart of the Fokker-Planck equation (Eq.~\eqref{eq:Fokker-Planck}): the latter is solved by the transition densities $\rho_t$ of the stochastic process $X$ that solves the former. In a similar sense, we want to establish a probabilistic counterpart of the Fokker-Planck-Birth-Death equation (Eq.~\eqref{eq:Fokker-Planck-Birth-Death}). To achieve this, we introduce the following interacting particle system that is also described in Ref.~\citenum{lu_accelerating_2019}.

We assume that there are $N$ particles with positions $x_1(t),\ldots, x_N(t)\in\R^d$ at time $t\ge0$. By
\begin{equation}\label{eq:empirical_measure}
    \mu_t^N=\frac 1 N \sum_{k=1}^N\delta_{x_k(t)}
\end{equation}
we denote the empirical measure of this $N$ particle system, i.e., $\mu_t^N$ puts a mass of $1/N$ at each of the $N$ particles' current positions. Since this is a singular measure and since we would like to be able to plug it into the birth-death term, we need to replace $\alpha_\pi(\cdot)$ with a smoothened approximation $\Lambda(\cdot)$, giving rise to an approximated Fokker-Planck-Birth-Death equation
\begin{equation}\label{eq:approxFPBD}
	\d_t\rho_t=L^* \rho_t -\tau_{\alpha}\Lambda (\rho_t) \rho_t.
\end{equation}
Now, we assume that each particle diffuses independently according to the overdamped Langevin dynamics defined in Eq.~\eqref{eq:SDE}. On top of these independent dynamics, the particles interact via the following mechanism. Each particle has an independent exponential clock that strikes with the configuration-dependent birth-death rate
\begin{equation}\label{eq:birth-death-rate}
    \tau_{\alpha}\,|\Lambda (\mu_t^N)(x_i(t))|.
\end{equation}
If the exponential clock for the $i$-th particle strikes at time $t$, then one of two things happens:
\begin{itemize}
	\item
	If $\Lambda (\mu_t^N)(x_i(t))>0$, then the $i$-th particle is killed and a particle chosen uniformly at random from the others is duplicated.
	\item
	If $\Lambda (\mu_t^N)(x_i(t))<0$, then the $i$-th particle is duplicated and a particle chosen uniformly at random from the others is killed.
\end{itemize}
Thus, the total number of particles is preserved. Alternatively, we can interpret this mechanism in the following way: in the first of the two cases above, the $i$-th particle jumps to the current position of a random other particle, and in the second case a random other particle jumps to the current position of the $i$-th particle.

Since $\Lambda$ is a smoothened version of $\alpha_\pi$, the term $\Lambda (\mu_t^N)(x_i(t))$ approximately takes the logarithmic difference between the current particle density at $x_i(t)$ and the equilibrium density $\pi(x_i(t))$, and then subtracts the average of the same quantity over all current particle positions $x_1(t),\ldots, x_N(t)$. Therefore, the birth-death mechanism has a tendency to kill particles in space regions that are currently very crowded relative to the energy, and to duplicate particles in the opposite situation. Hence, we can expect this birth-death mechanism to help distribute the particles according to $\pi$ and thus speed up the convergence of $\mu_t^N$ to $\pi$ in comparison to a system without birth-death events.

We will now discuss three different choices of the approximation $\Lambda$, all of which feature a centered Gaussian kernel
\begin{equation}\label{eq:gaussian}
    K_{\boldsymbol\Sigma}(x)=\frac{1}{(2 \pi)^{d/2}\lvert\boldsymbol{\Sigma}\rvert^{1/2}} \exp\argleft(-\frac{x^\top \boldsymbol{\Sigma}^{-1}x}{2}\argright), \quad x\in\R^d,
\end{equation}
where $\boldsymbol{\Sigma}\in\R^{d\times d}$ is a positive definite invertible covariance matrix and $\lvert\boldsymbol{\Sigma}\rvert$ denotes its determinant. Since $\boldsymbol\Sigma$ is usually fixed, we will often just write $K=K_{\boldsymbol\Sigma}$. For our simulations, we are mostly interested in a diagonal covariance matrix, $\Sigma_{ij} = \delta_{ij} \sigma_{i}^2$ with $\sigma_{i}>0$, where $\delta_{ij}$ is the Kronecker delta. In this case, $\boldsymbol{\sigma}=(\sigma_{1},\ldots,\sigma_{d})\in\R^d$ will be referred to as the (vector of) bandwidths of the smoothing kernel. We write
\begin{equation}\label{eq:convolution}
    K*f(x)=\int K(x-y)f(y)dy
\end{equation}
for the convolution of $K$ with a (generalized) function $f$. In particular, for the empirical distribution $\mu_t^N$ from Eq.~\eqref{eq:empirical_measure}, $K*\mu_t^N$ can be thought of as a kernel density estimation~\cite{silverman_density_1998}.

In Ref.~\citenum{lu_accelerating_2019}, the authors use the approximation
\begin{align}
    \Lambda^0 (f)&=\alpha_\pi (K*f,f)
    \nonumber \\
    &=\log\frac{K*f(x)}{\pi(x)}-\int\log\left(\frac{K*f(y)}{\pi(y)}\right)f(y) \diff y.
\end{align}
This choice has one crucial shortcoming, as $\Lambda^0 (\pi)\neq0$ and hence $\pi$ is not a stationary solution to the corresponding approximation in Eq.~\eqref{eq:approxFPBD} of the Fokker-Planck-Birth-Death equation. In other words, this approximation changes the equilibrium and hence using it as the basis for a sampling algorithm leads to sampling the wrong probability distribution.

One way to solve this problem is offered by the alternate approximation
\begin{equation}\label{eq:correction_add}
    \Lambda^{\mathrm{ad}} (f)
    =\Lambda^0 (f)-\Lambda^0 (\pi)
\end{equation}
that adds a correction term to $\Lambda^0$. Clearly, $\Lambda^{\mathrm{ad}} (\pi)=0$. Unfortunately however, this additive correction is not particularly convenient for the mathematical analysis (see e.g., the comments after Theorem 1).
Hence, we consider the alternative ``multiplicative'' correction
\begin{align}
\label{eq:correction_mu}
    \Lambda^{\mathrm{mu}} (f)&=\alpha_{K*\pi} (K*f,f)
    \nonumber \\
    &=\log\frac{K*f(x)}{K*\pi(x)}-\int\log\left(\frac{K*f(y)}{K*\pi(y)}\right)f(y) \diff y,
\end{align}
where the equilibrium distribution $\pi$ is also convoluted with a Gaussian kernel. Here, again, $\Lambda^{\mathrm{mu}} (\pi)=0$. Unless explicitly stated otherwise, we will usually work with $\Lambda=\Lambda^{\mathrm{mu}}$.

Note that if we formally set $\boldsymbol\Sigma=0$ and interpret $K_0$ as the dirac delta $\delta_0$, then all of the approximations $\Lambda^0$, $\Lambda^{\mathrm{mu}}$, $\Lambda^{\mathrm{ad}}$ coincide with the original birth-death term $\alpha_\pi$. We establish two important theoretical properties of these interacting particle systems that explain and complement our findings in the subsequent sections on practical applications. First, we show that for fixed times $t>0$, the empirical measure $\mu_t^N$ converges weakly to the solution $\rho_t$ of the approximation of the Fokker-Planck-Birth-Death equation (Eq.~\eqref{eq:approxFPBD}) if the number $N$ of particles tends to infinity (Theorem 1). In particular, this gives proper meaning to the idea that this interacting particle system is the probabilistic counter-part of the Fokker-Planck-Birth-Death equation. Second, we present reasonable assumptions under which $\rho_t$ converges (exponentially fast) to $\pi$, as time $t$ goes to infinity (Theorem 2). This rough summary is enough to understand the applications below, so the following section may be skipped on first reading.

\textbf{Remark.} In a recent talk~\cite{YulongLu_Talk2021}, one of the authors of Ref.~\citenum{lu_accelerating_2019} presented the approximation
\begin{align}
    \Lambda^{0+} (f)=\log K*\frac{f}{\pi}(x)-\int\log\left(K*\frac{f}{\pi}(y)\right)f(y) \diff y,
\end{align}
which solves the main issue of $\Lambda^{0}$, as $\Lambda^{0+}(\pi)=0$ Furthermore, it still satisfies Eq.~\eqref{eq:Lambdacentered}, so Theorem 1 also holds with $\Lambda=\Lambda^{0+}$.

\subsection{Theoretical results}

\textbf{Theorem 1.} Let $\rho_t$ be the solution to Eq.~\eqref{eq:approxFPBD} with $\Lambda\in\{\Lambda^0,\Lambda^{\mathrm{mu}}\}$ and assume that $\mu_0^N$ converges weakly to the probability measure with density $\rho_0$ for $N\to\infty$.
Then for all $t>0$, the empirical measure $\mu_t^N$ of the interacting particle system converges weakly to the probability measure with density $\rho_t$ for $N\to\infty$.

\medskip

\textbf{Non-rigorous proof.} For the sake of notational simplicity and without loss of generality, we assume that $D=\tau_\alpha=1$. Let $(x_n)_{n\in\N}\subset\R^d$. For any $N\in\N$, write
\begin{equation}\label{eq:mu^N}
    \mu_x^N=\frac 1 N \sum_{k=1}^N\delta_{x_k}
\end{equation}
and let $\mu_x$ denote the formal limit of $\mu_x^N$ for the number $N$ of particles going to infinity. The main idea of this proof is to show that for any smooth functional $\Psi$ mapping a probability measure on $\R^d$ to an element of $\R^d$, we have convergence of $(\cL_N\Psi)(\mu_x^N)$ to $(\cL\Psi)(\mu_x)$, where $\cL_N$ is the generator of the measure valued Markov process given by the empirical measure $\mu_t$ of the interacting particle system (see Eq.~\eqref{eq:empirical_measure}), and $\cL$
corresponds to how the right hand side of Eq.~\eqref{eq:approxFPBD} acts on the functional $\Psi$.

If $\mu_x^N$ is the current configuration of the system and the clock strikes for the $i$-th particle, an index $j\in\{1,\ldots,N\}\setminus\{i\}$ is chosen uniformly at random, and the configuration changes to
\begin{equation}
    \mu_x^N(x_i \to x_j):=\mu_x^N+\frac 1 N \sign\argleft(\Lambda (\mu_x^N)(x_i)\argright) (\delta_{x_j}-\delta_{x_i}).
\end{equation}
This happens at rate $\Lambda (\mu_x^N)(x_i)$, and in between these birth-death events, each particle diffuses independently according to Eq.~\eqref{eq:SDE}. Therefore, the infinitesimal generator $\cL_N$ is given by
\begin{align}\label{eq:proofgenerator}
    \begin{split}
        &(\cL_N\Psi)(\mu_x^N) \\
        =&\frac 1 N \sum_{i=1}^N\left(\Delta \Psi'_{\mu_x^N}(x_i)+\nabla \Psi'_{\mu_x^N}(x_i)\cdot\nabla\log\pi(x_i) \right)\\
        & +\frac 1 N \sum_{i,j=1}^N |\Lambda (\mu_x^N)(x_i))|\left(\Psi(\mu_x^N(x_i \to x_j))-\Psi(\mu_x^N)\right)
    \end{split}
\end{align}
for any smooth functional $\Psi$, where $\Psi'_{\cdot}$  denotes its functional derivative, i.e.,
\begin{equation}\label{eq:functionalderivative}
    \int \Psi'_{\nu_0}(x) \nu(\diff x)=\lim_{\eps\to0}\frac{\Psi(\nu_0+\eps\nu)-\Psi(\nu_0)}{\eps}
\end{equation}
for any probability measure $\nu_0$ and any measure $\nu$ with $\int x \nu(\diff x)=0$ (i.e., $\nu$ is centered). With the help of Eq.~\eqref{eq:mu^N}, we can rewrite Eq.~\eqref{eq:proofgenerator} as
\begin{align}
    \begin{split}
        &(\cL_N\Psi)(\mu_x^N) \\
        =&\int\left(\Delta \Psi'_{\mu_x^N}(y)+\nabla \Psi'_{\mu_x^N}(y)\cdot\nabla\log\pi(y) \right)\mu_x^N(\diff y)\\
        & + N \int\int |\Lambda (\mu_x^N)(y))|\left(\Psi(\mu_x^N(y \to z))-\Psi(\mu_x^N)\right) \\
        &\hspace{5.7cm}\mu_x^N(\diff y)\mu_x^N(\diff z).
    \end{split}
\end{align}
The formal limit of the first summand is
\begin{equation}
    \int\left(\Delta
    \Psi'_{\mu_x}(y)+\nabla \Psi'_{\mu_x}(y)\cdot\nabla\log\pi(y) \right)\mu_x(\diff y).
\end{equation}
For the second summand, we note that by Eq.~\eqref{eq:functionalderivative} with $\eps=\frac1N$ and $\nu=\sign(\Lambda (\mu_x^N)(y)) (\delta_{y}-\delta_{z})$, we have
\begin{align}
    \begin{split}
        &\Psi(\mu_x^N(y \to z))-\Psi(\mu_x^N) \\
        \approx& \frac 1 N \int \Psi'_{\mu_x^N}(u)  \sign(\Lambda (\mu_x^N)(y)) (\delta_{z}-\delta_{y})(\diff u)
    \end{split}
\end{align}
for $N\to\infty$. Therefore, the second summand formally converges to
\begin{align}
    \begin{split}
        &\int\int\int \Lambda (\mu_x)(y)\Psi'_{\mu_x}(u)  (\delta_{z}-\delta_{y})(\diff u)\mu_x(\diff y)\mu_x(\diff z) \\
        =&\int \left(\int \Lambda (\mu_x)(z)\mu_x(\diff z)-\Lambda(\mu_x)(y)\right) \Psi'_{\mu_x}(y)\mu_x(\diff y),
    \end{split}
\end{align}
and hence $\cL_N\Psi(\mu_x^N)$ converges to
\begin{align}\label{eq:limitgenerator}
    \begin{split}
        & \quad (\cL\Psi)(\mu_x) \\
        &=\int\left(\Delta
        \Psi'_{\mu_x}(y)+\nabla \Psi'_{\mu_x}(y)\cdot\nabla\log\pi(y) \right)\mu_x(\diff y)\\
        & \qquad + \int \left(\int \Lambda (\mu_x)(z)\mu_x(\diff z)-\Lambda(\mu_x)(y)\right) \\
        & \hspace{5cm} \Psi'_{\mu_x}(y)\mu_x(\diff y).
    \end{split}
\end{align}
Since for $\Lambda\in\{\Lambda^0,\Lambda^{\mathrm{mu}}\}$ we have
\begin{equation}\label{eq:Lambdacentered}
    \int \Lambda (\mu_x)(z)\mu_x(\diff z)=0,
\end{equation}
the formal limit generator $(\cL\Psi)(\mu_x)$ from Eq.~\eqref{eq:limitgenerator} corresponds to Eq.~\eqref{eq:approxFPBD} with the respective choice of $\Lambda\in\{\Lambda^0,\Lambda^{\mathrm{mu}}\}$.
\hfill$\square$

\medskip

Our proof uses the same arguments as the proof of Proposition 5.1 in Ref.~\citenum{lu_accelerating_2019} where the case $\Lambda=\Lambda^0$ was already treated. Note that up until Eq.~\eqref{eq:limitgenerator}, the proof works for any choice of $\Lambda$. However, if we use $\Lambda=\Lambda^{\mathrm{ad}}$, we have to be a little more careful, since then $\int \Lambda^{\mathrm{ad}} (\mu_x)(y)\mu_x(\diff y)$ does not vanish in general. The limit generator $(\cL\Psi)(\mu_x)$ from Eq.~\eqref{eq:limitgenerator} then corresponds to Eq.~\eqref{eq:approxFPBD} with
\begin{equation}
    \Lambda(f)=\Lambda^{\mathrm{ad}}(f)-\int \Lambda^{\mathrm{ad}}(f)f\diff x.
\end{equation}
Unfortunately, it is currently unclear to us how one could say anything about the corresponding stationary solution.

\medskip

In order to quantify the distance between $\rho_t$ and $\pi$, we will use the Kullback--Leibler divergence
\begin{equation}\label{eq:kl_div}
    D_{\mathrm{KL}}(\rho_t|\pi)=\int\log\left(\frac{\rho_t}{\pi}\right)\rho_t \diff x,
\end{equation}
even though it is not a metric in the mathematical sense (as it is not symmetric and also violates the triangle inequality). However, it can be related to an actual metric, as Pinsker's inequality shows that the property $D_{\mathrm{KL}}(\rho_t|\pi)\to0$ is stronger than convergence of $\rho_t$ to $\pi$ with respect to the total variation distance.

We will also use the relative Fisher information
\begin{equation}
    \cI(\rho_t|\pi)=\int\left|\nabla\log\frac{\rho_t}{\pi}\right|^2\rho_t \diff x.
\end{equation}

The following Theorem contains Theorem 3.2 of Ref.~\citenum{lu_accelerating_2019} as a special case, as for $K=\delta_0$ (i.e., $\Lambda=\alpha_\pi$), the second assumption trivially holds with $\lambda'=0$.

\medskip

\textbf{Theorem 2.} Let $\rho_t$ the solution to Eq.~\eqref{eq:approxFPBD} with $\Lambda=\Lambda^{\mathrm{mu}}$ and assume that the following conditions hold.
\begin{itemize}
	\item
	There is a $\lambda>0$ such that the log-Sobolev-inequality
	\begin{equation}\label{eq:logsobolev}
        D_{\mathrm{KL}}(f|\pi)\le \frac{1}{\lambda}\cI(f|\pi)
    \end{equation}
	holds for all probability densities $f$ on $\R^d$.
	\item
	There is a $\lambda'>-D\lambda/\tau_\alpha$ such that
	\begin{equation}\label{eq:covariancecondition}                             \mathrm{Cov}_{\rho_t}\left(\log\frac{\rho_t}{\pi},\log\frac{K*\rho_t}{K*\pi} \right)\ge \lambda' D_{\mathrm{KL}}(\rho_t|\pi)
    \end{equation}
	for all $t>0$, where
	\begin{align}
        \mathrm{Cov}_{\rho}\left(f,g \right)
         =\int fg\rho \diff x-\int f\rho \diff x \int g\rho \diff x.
    \end{align}
\end{itemize}
Then
\begin{equation}\label{eq:Theorem2convergence}
    D_{\mathrm{KL}}(\rho_t|\pi)\le D_{\mathrm{KL}}(\rho_0|\pi)e^{-t(D\lambda+\tau_\alpha\lambda')}\xrightarrow{t\to\infty}0.
\end{equation}

\medskip

\textbf{Proof.} Since $\rho_t$ is a smooth probability density, we have
\begin{equation}
    \int(\d_t\rho_t)\diff x=\d_t\int\rho_t\diff x=0
\end{equation}
and hence
\begin{align}
\begin{split}
    \d_t D_{\mathrm{KL}}(\rho_t|\pi)
    & =\int\d_t\left(\log\left(\frac{\rho_t}{\pi}\right)\rho_t\right) \diff x \\
    & =\int(\d_t\rho_t)\left(\log\frac{\rho_t}{\pi}+1\right) \diff x \\
    & =\int(\d_t\rho_t)\log\frac{\rho_t}{\pi}\diff x.
\end{split}
\end{align}
Plugging in Eq.~\eqref{eq:approxFPBD} and using integration by parts yields
\begin{align}
\begin{split}
\label{eq:KL_derivative}
    & \quad \d_t D_{\mathrm{KL}}(\rho_t|\pi) \\
    & =\int\left(D\nabla\cdot\left(\rho_t \nabla \log\frac{\rho_t}{\pi}\right)-\tau_\alpha\Lambda^{\mathrm{mu}}(\rho_t) \rho_t\right)\log\frac{\rho_t}{\pi}\diff x \\
    & = -D\cI(\rho_t|\pi)-\tau_\alpha\mathrm{Cov}_{\rho_t}\left(\log\frac{\rho_t}{\pi},\log\frac{K*\rho_t}{K*\pi} \right).
\end{split}
\end{align}
Eq.~\eqref{eq:KL_derivative} together with our assumptions Eq.~\eqref{eq:logsobolev} and Eq.~\eqref{eq:covariancecondition} implies
\begin{equation}
    \d_t D_{\mathrm{KL}}(\rho_t|\pi)\le-(D\lambda+\tau_\alpha\lambda')D_{\mathrm{KL}}(\rho_t|\pi).
\end{equation}
The claim now follows from Gronwall's Lemma.
\hfill$\square$

\medskip

Note that the parameter $\lambda>0$ in the log-Sobolev-inequality Eq.~\eqref{eq:logsobolev} is present explicitly in the convergence rate in Eq.~\eqref{eq:Theorem2convergence} and, of course, depends crucially on the potential $U$.
This dependence can be described via the Eyring–Kramers formula for log-Sobolev-inequalities (Corollary 2.17 in Ref.~\citenum{log-sobolev}).
For a double-well potential, $\lambda$ decreases exponentially with respect to the height of the energy barrier (compare Corollary 2.18 in Ref.~\citenum{log-sobolev})

The covariance condition in Eq.~\eqref{eq:covariancecondition} may be difficult to establish in practice, but let us present a rough idea why it is plausible. If $\rho_t$ and $\pi$ are sufficiently smooth, the amount to which they change after applying the smoothing kernel $K$ is bounded uniformly in time and space by some constant times $|\boldsymbol\Sigma|$. If we pretend that $\pi$ and $\rho_t$ simply vanish entirely in very high energy regions, we can then argue that
\begin{align}\label{eq:covbound1}
    \begin{split}
    &\int\left(\log\frac{\rho_t}{\pi}\right)\left(\log\frac{K*\rho_t}{K*\pi}\right)\rho_t\diff x \\
    \ge\;& \int\left(\log\frac{\rho_t}{\pi}\right)\left(\log\frac{K*\rho_t}{K*\pi}-\log\frac{\rho_t}{\pi}\right)\rho_t\diff x \\
    \approx\;& \eps(\boldsymbol\Sigma) \cdot D_{\mathrm{KL}}(\rho_t|\pi),
    \end{split}
\end{align}
where $|\eps(\boldsymbol\Sigma)|$ is small for $\boldsymbol\Sigma\to0$. If our initial condition $\rho_0$ is not too far off, it is also reasonable to expect that $\rho_t\le C\pi$ for some $C\in(1,\infty)$. If this holds, we can also estimate
\begin{equation}\label{eq:covbound2}
    \int\log\frac{\rho_t}{\pi}\rho_t\diff x\int\log\frac{K*\rho_t}{K*\pi}\rho_t\diff x\le (\log C) D_{\mathrm{KL}}(\rho_t|\pi).
\end{equation}
If $\boldsymbol\Sigma$ and $C$ can be chosen suitably, combining Eq.~\eqref{eq:covbound1} and Eq.~\eqref{eq:covbound2} yields a version of the covariance condition in Eq.~\eqref{eq:covariancecondition}.

In Section S-I of the Supplemental Material (SM)~\cite{supplemental_material}, we show empirically that the covariance condition in Eq.~\eqref{eq:covariancecondition} is satisfied for an exemplary simulation.
There we find $\lambda'$ to have a non-negative lower bound, so we can plausibly assume Eq.~\eqref{eq:covariancecondition} to hold with $\lambda' = 0$, as is also the case when no smoothing kernel is applied at all.

\medskip

The following Lemma gives meaning to the notion that increasing the bandwidth of the smoothing kernel corresponds to turning off the birth-death-mechanism.

\medskip

\textbf{Lemma 1.} If $f$ is a probability density function on $\R^d$, then
\begin{equation}\label{eq:lemma1}
    \frac{K_{\boldsymbol\Sigma} * f}{K_{\boldsymbol\Sigma}}\xrightarrow{|\boldsymbol\Sigma|\to\infty} 1
\end{equation}
pointwise on $\R^d$. In particular, if
\begin{equation}\label{eq:lemma1.5}
    C^{-1}\pi\le f\le C\pi
\end{equation}
for some $C>1$, then
\begin{equation}\label{eq:lemma2}
    \Lambda^{\mathrm{mu}} (f) \xrightarrow{|\boldsymbol\Sigma|\to\infty} 0
\end{equation}
pointwise on $\R^d$.

\medskip

\textbf{Proof.} First, Eq.~\eqref{eq:lemma1} can be shown by a straight forward calculation and using dominated convergence. Then,
\begin{equation}\label{eq:lemmaproof}
    \frac{K_{\boldsymbol\Sigma} * f}{K_{\boldsymbol\Sigma}*\pi}=\frac{K_{\boldsymbol\Sigma} * f}{K_{\boldsymbol\Sigma}}\frac{K_{\boldsymbol\Sigma}}{K_{\boldsymbol\Sigma} * \pi}\xrightarrow{|\boldsymbol\Sigma|\to\infty} 1,
\end{equation}
so that Eq.~\eqref{eq:lemma2} follows from Eq.~\eqref{eq:lemma1.5} and another application of dominated convergence. \hfill$\square$

\medskip

Note that
\begin{align}
    \begin{split}
    \Lambda^{\mathrm{ad}}_\pi (f)&=\log \frac{K_{\boldsymbol\Sigma} * f}{K_{\boldsymbol\Sigma}*\pi}
	-\int\left(\log\frac{K_{\boldsymbol\Sigma} * f}{\pi}\right)f \diff x \\
	& \quad +\int\left(\log\frac{K_{\boldsymbol\Sigma} * \pi}{\pi}\right)\pi \diff x,
	\end{split}
\end{align}
where Eq.~\eqref{eq:lemmaproof} implies that the first term goes to zero for $|\boldsymbol\Sigma|\to\infty$, while the remaining terms do not depend on the position. Hence, even though the birth-death mechanism may not be turned off entirely in the limit, it no longer distinguishes between the different particles.

\medskip

\subsection{The underdamped Langevin dynamics case}\label{sec:theory_underdamped_langevin}
Examining the behavior of the birth-death term for overdamped Langevin dynamics makes it feasible to mathematically prove convergence to the right distribution. However, for physical and chemical systems we often employ more general dynamics that take into account inertia and thus have to track not only the particle's position $x(t)$, but also its momentum $p(t)$. This is described by the underdamped Langevin equations
\begin{align}
  \label{eq:underdamped_langevin1}
  \diff x(t) &= \frac{p(t)}{m} \diff t, \\
  \label{eq:underdamped_langevin2}
  \diff p(t) &= - \nabla U(x(t))\diff t - \gamma p(t) \diff t + \sqrt{\frac{2 m \gamma}{\beta} }\diff W(t),
\end{align}
where $m$ denotes the particle mass, $\gamma$ is a friction constant, and $\beta^{-1}=k_{\mathrm{B}}T$ is the thermal energy as before. Note that setting $\diff p(t)=0$ in Eq.~\eqref{eq:underdamped_langevin2}, plugging in Eq.~\eqref{eq:underdamped_langevin1}, and rearranging the terms yields Eq.~\eqref{eq:SDE} with $D=(m\gamma\beta)^{-1}$.

The solution $(X,P)=(x(t),p(t))_{t\ge0}$ of Eqs.~\eqref{eq:underdamped_langevin1} and \eqref{eq:underdamped_langevin2} is a $2d$-dimensional Markov process. It possesses a unique invariant distribution whose marginal with respect to the position $x$ coincides with $\pi$, since in equilibrium, position and momentum become independent. In order to introduce a birth-death mechanism to an ensemble of $N$ particles diffusing according to Eqs.~\eqref{eq:underdamped_langevin1} and \eqref{eq:underdamped_langevin2}, we follow the same approach as in Section \ref{sec:theory_algorithm} and still use the same birth-death term $\Lambda (\mu_t^N)(x_i(t))$ that depends only on the positions and ignores the momenta. When the $i$-th particle is killed (or duplicated), the entire tuple $(x_i,p_i)$ is killed (or duplicated).

While we do not present any theory for the resulting interacting particle system, we investigate it empirically in simulations in Section \ref{sec:underdampedlangevinsimulations} and verify that our algorithm can also successfully be used to sample $\pi$ in the underdamped Langevin case.
Additionally we provide an analysis of the momentum distribution and correlation for an exemplary simulation in Section~S-II of the SM, where we conclude that the chosen approach does not result in deviations for the momentum distribution and equilibration and is therefore justified for the presented system.

\section{Implementation}\label{sec:implementation}

In the particle based view, we can explicitly give the formula for the birth-death term. We can rewrite Eq.~\eqref{eq:correction_mu} as
\begin{align}\label{eq:lambda_mu_explicit}
    \begin{split}
        \Lambda^{\mathrm{mu}} (f)(x)&= \log K*f(x)-\log K*\pi(x) \\
            & \quad -\int\big(\log K*f(y)-\log K*\pi(y) \big)f(y) \diff y
    \end{split}
\end{align}
for all $x\in\R^d$. Since
\begin{equation}
    K*\delta_y(x)=\int K(x-z)\delta_y(z)dz=K(x-y)
\end{equation}
for any $x,y\in\R^d$ and since convolution is a linear operation, we can easily plug the empirical measure
\begin{equation}
    \mu_t^N=\frac 1 N \sum_{k=1}^N\delta_{x_k(t)}
\end{equation}
into Eq.~\eqref{eq:lambda_mu_explicit}. Dropping the time dependence of the particle positions $x_1(t),\ldots,x_N(t)$ for notational convenience, we find that
\begin{equation}\label{eq:lambda_particle}
    \begin{aligned}
        \Lambda^{\mathrm{mu}} (\mu_t^N)(x_i) = \log\argleft(\frac{1}{N} \sum_{j=1}^N K(x_i - x_j)\argright) - \log(K * \pi(x_i))\\
        - \frac{1}{N}\sum_{k=1}^N\left[\log\argleft(\frac{1}{N} \sum_{j=1}^N K(x_k - x_j)\argright) - \log(K * \pi(x_k))\right].
    \end{aligned}
\end{equation}
In the sequel, we will simply write
\begin{equation}
    \Lambda_i:=\Lambda^{\mathrm{mu}}(\mu_t^N)(x_i).
\end{equation}
We will always present results with $\Lambda=\Lambda^{\mathrm{mu}}$ in the following, although we show in the supplemental material~\cite{supplemental_material} that similar results are obtained with $\Lambda=\Lambda^{\mathrm{ad}}$.

In our simulations we choose the covariance matrices of the Gaussian kernel as diagonal,  $\Sigma_{ij} = \delta_{ij} \sigma_{i}^2$ with the bandwidths $\boldsymbol{\sigma} = (\sigma_{1},\ldots,\sigma_{d})$ where $\sigma_{i}>0$. Then Eq.~\eqref{eq:gaussian} turns into
\begin{equation}
    K(x)=K_{\boldsymbol{\sigma}}(x) = \frac{1}{ (2\pi)^{d/2} \prod_{i=1}^{d} \sigma_{i}}
    \exp\argleft(-
    \sum_{i=1}^{d} \left(\frac{x^{(i)}}{\sqrt{2}\sigma_{i}}\right)^2\argright)
\end{equation}
where the sum goes over the $d$ spatial dimensions of the state $x=(x^{(1)},\ldots,x^{(d)})\in\R^d$ of an individual particle.

Our algorithm mostly follows Algorithm 1 of Ref.~\citenum{lu_accelerating_2019} but with modifications to reduce the computational effort and the previously mentioned changes to the calculation of the approximate birth-death term $\Lambda$.
First, while the original algorithm proposed to attempt birth-death events after every Langevin step, we do so only every $M$ steps.
This results in calculating the $\Lambda$ values, which require the computationally involving density estimate $\rho_t$, less often.
This has to be taken into account for the exponential clock: if $q_i$ denotes the probability that the clock of the $i$-th particle strikes after $M$ steps, then these birth-death probabilities become
\begin{equation}\label{eq:bd_probs}
    q_i=1-\exp\left(-\tau_{\alpha}\lvert \Lambda_i \rvert M \theta \right),
\end{equation}
where $\theta$ denotes the Langevin time step.
We will test if this results in deviation in the sampling.
As in Ref.~\citenum{lu_accelerating_2019}, we will use $\tau_{\alpha} = 1$ for the following applications but present a short discussion about the parameter in Section S-III in the SM~\cite{supplemental_material}.

Second, the original algorithm of Ref.~\citenum{lu_accelerating_2019} iterates over the particles, where it individually calculates the birth-death probability and executes accepted events immediately.
Therefore, the values $\Lambda_i$ have to be calculated for each particle individually, or at least recalculated from the new positions after each accepted birth-death event.
For efficiency, we instead choose to calculate all birth-death rates $\tau_{\alpha}\Lambda_i$ from the positions only once before the birth-death step.
Only the order in which the birth-death events are applied is randomized.
No disadvantages could be found from this approach as long as the probabilities of birth-death events remain low, as will be investigated further in the following.

\begin{algorithm}\label{alg:algorithm}
    \caption{Birth-death augmented Langevin dynamics}
    \DontPrintSemicolon
    \KwIn{
    \begin{itemize}
        \item Potential $U$ (and temperature $T$) corresponding to the equilibrium distribution $\pi$
        \item Langevin solver $L(X,P,U,\theta)$ with corresponding parameters
        \item Calculation rule for smoothed birth-death term $\Lambda$ using Gaussian kernel $K$ with bandwidths $\boldsymbol{\sigma}$
        \item Rate factor $\tau_\alpha$
        \item Langevin time step $\theta$
        \item Number of Langevin steps $J$
        \item Number of Langevin steps between birth-death attempts $M$
        \item $N$ particles with initial positions $X=\{x_i\}_{i=1}^N$ and momenta $P=\{p_i\}_{i=1}^N$
    \end{itemize}
    }
    \KwOut{
    \begin{itemize}
        \item Set of particles whose empirical measure approximates $\pi$
    \end{itemize}
    }
    \For{$t\leftarrow 1$ \KwTo $J$}{
        update $X$ and $P$ by Langevin solver $L(X,P,U,\theta)$\;
        \If{$(t \mod M) = 0$}{
            Calculate $\Lambda$ for all particles\;
            Draw $N$ independent random numbers $\{r_i\}_{i=1}^N$ uniformly from $[0,1)$\;
            Make list $\zeta$ of indices $i$ for which $r_i \leq q_i= 1-\exp\left(- \tau_\alpha \lvert \Lambda_i \rvert M \theta \right)$\;
            Shuffle $\zeta$ randomly\;
            \ForEach{$i \in \zeta$ \footnote{In the foreach loop, we skip over all particles that were already killed randomly by a previous duplication event during the same birth-death step. This avoids duplicating the new position of a killed particle that was not actually considered for the event probability.}}{
                Select particle $j$ uniformly from all other particles\;
                \uIf{$\Lambda_i > 0$}{
                    $x_i \leftarrow x_j$; $p_i \leftarrow p_j$\;
                    }
                \ElseIf{$\Lambda_i < 0$}{
                    $x_j \leftarrow x_i$; $p_j \leftarrow p_i$\;
                }
            }
        }
    }
\end{algorithm}

The algorithm was implemented in a custom Python code together with Langevin solvers (i.e., integrators), and has been made available to the community~\cite{bdld_software}.
Version v0.3.1 was used for all calculations in the following.
We employ the Euler-Maruyama scheme~\cite{kloeden_numerical_1999} for overdamped Langevin dynamics and the Bussi-Parinello scheme~\cite{bussi_accurate_2007} for the underdamped Langevin dynamics case.
The input files and data supporting the results of this paper are openly available at Zenodo~\cite{pampel_dataset_2022} (DOI: \href{https://zenodo.org/record/5873264}{\texttt{10.5281/zenodo.5873264}}).

\section{Applications}\label{sec:applications}

As test cases, we choose to simulate the movement of sets of particles in artificial potentials that emulate prototypical energy landscapes. The focus is on rare event systems, where the energy landscapes are characterized by metastable states separated by high kinetic barriers (i.e., much higher than the thermal energy $k_{\mathrm{B}}T$) that hinder transitions between metastable states. Such rare event energy landscapes are common in the physical sciences so we foresee many applications that can benefit from this birth-death method.

\subsection{Comparison of approximations $\Lambda$}\label{sec:appl_approximation_comparison}
\begin{figure*}
    \includegraphics{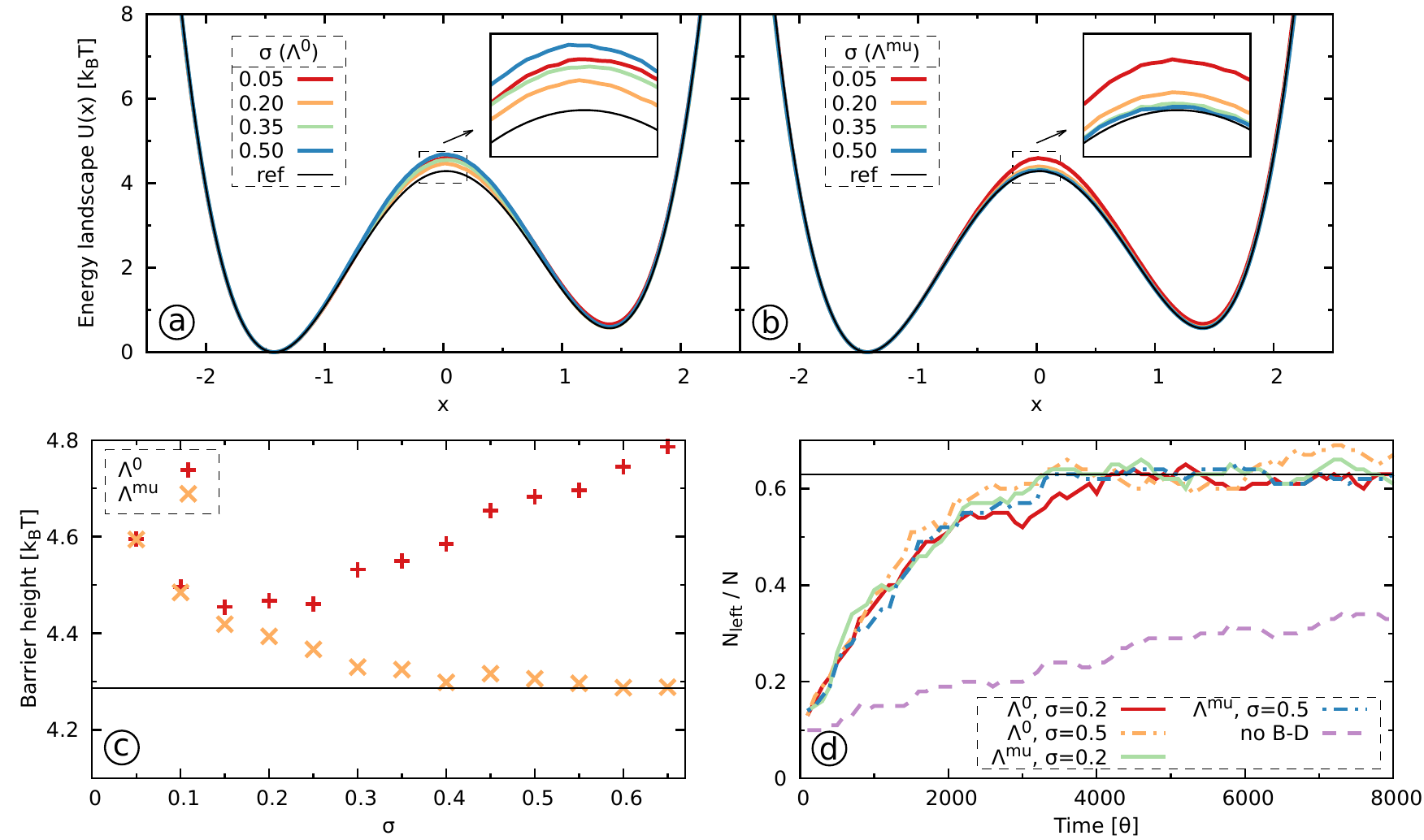}
    \caption{
        Estimates of the energy landscape for the potential given in Eq.~\eqref{eq:double_well} obtained from sampling using histogramming for the different approximations to the birth-death term:
        (a) The original proposal $\Lambda^0$.
        (b) The new proposal $\Lambda^{\mathrm{mu}}$. Both show results for different values of the kernel width $\sigma$ (colored) as well as the reference (black). The small inset shows a magnification of the barrier region.
        (c) The height of the barrier going from the left minimum to the right minimum estimated from the energy landscape as a function of the kernel width. The reference value is given as a black horizontal line.
        (d) Fraction of particles in the left state as a function of simulation time. The black horizontal line is the expected equilibrium value.
        Shown are results from two exemplary simulations with different kernel widths for both approximations ($\sigma =0.2$ as solid line, $\sigma = 0.5$ as lines with dashes and dots), as well as from a simulation without birth-death events (dashed line). Only the first 8,000 steps of the simulations are displayed.
    \label{fig:correction}}
\end{figure*}

To be able to show the effects of the different approximations $\Lambda$ and other parameters, we start with a system with a moderate barrier height such that transitions are also observed within moderate simulation time by pure Langevin dynamics.
We choose a one-dimensional double-well energy landscape that is described by the mathematical expression
\begin{equation}\label{eq:double_well}
  U(x) = x^4 - 4x^2 + 0.2x + C,
\end{equation}
where $C$ is a constant so that $\min U(x) = 0$ to align it for easier comparison.
A plot of this potential can be seen as a black line in Fig.~\ref{fig:correction}(a).
It features two metastable states that have local minima at $x_L\approx-1.4$ and $x_R\approx1.4$, and are separated by a barrier of $4.285\,k_\mathrm{B}T$ that is centered at the local maximum at $x_0\approx0$. The metastable state at $x_L$ on the left is lower in energy and thus has higher probability.  In equilibrium, the probability of a particle being in the basin of attraction $B_L=(-\infty,x_0)$ for the left metastable state is $\pi(B_L)=\int_{-\infty}^{x_0} \pi(x) \diff x\approx0.63$. Similarly, the probability for the higher energy metastable state at $x_R$ on the right is $\pi(B_R)=\pi((x_0,\infty))\approx0.37$.

First, we test the influence of the newly proposed birth-death term $\Lambda^{\mathrm{mu}}$ on the sampling compared to the previously proposed birth-death term  $\Lambda^0$ from Ref.~\citenum{lu_accelerating_2019}.
We use $N=100$ particles and choose an initial distribution far from equilibrium: only 10 particles are placed in the more likely left state at $x_L$, while the remaining 90 particles start in the less likely right state at $x_R$.
We use the overdamped Langevin solver with a time step of $\theta = 0.001$ and set $D=1$ (see Section S-IX of the SM~\cite{supplemental_material} for a short discussion on the choice of the Langevin time step). We run the simulations for 2,000,000 steps at $T=1$ so that the thermal energy is $k_\mathrm{B}T=\beta^{-1}=1$ (we use natural units such that $k_\mathrm{B}=1)$. The number of steps between birth-death attempts is fixed to $M=100$, while we investigate both $\Lambda^0$ and $\Lambda^{\mathrm{mu}}$ with different kernel bandwidths $\sigma$.

We assess the correctness of the sampling by obtaining estimates of the energy landscape from the simulations via histogramming.
We bin all particle positions into a suitable histogram $H$ and at the end of the simulations we calculate the estimated energy landscape $\tilde{U}(x)$ via
\begin{equation}
    \tilde{U}(x) = -\beta^{-1} \log H(x) + \tilde{C},
\end{equation}
where we choose the constant $\tilde{C}$ such that $\min \tilde{U}(x) = 0$. When constructing the histograms, we always omit the first $10^5$ steps.

We show results for the different kernel bandwidths $\sigma$ in Fig.~\ref{fig:correction}(a,b).
We can observe that all simulations sample the basins and the lower regions of the energy landscapes correctly as indicated by the good agreement with the reference. However, we observe deviation from the reference energy landscape in the barrier region for some simulations, as can be seen in the insets in Fig.~\ref{fig:correction}(a,b). In particular, the deviation is larger for the birth-death term $\Lambda^0$ from Ref.~\citenum{lu_accelerating_2019}. To quantify the deviation, we calculate from the estimated energy landscapes the height of the barrier going from the left minimum to the right minimum. The estimated barrier heights for both birth-death terms are shown in Fig.~\ref{fig:correction}(c) as a function of the kernel width $\sigma$ together with the reference value.
We observe that for the original birth-death term  $\Lambda^0$, the barrier height is always overestimated as compared to the reference value. In other words, the original birth-death term $\Lambda^0$ leads to an undersampling of the barrier region.
On the other hand, the results for our new birth-death term $\Lambda^{\mathrm{mu}}$ are much better and we only observe an undersampling if the kernel bandwidth is very small. The reason for this effect is likely the kernel density estimate in Eq.~\eqref{eq:lambda_particle}, as a too small bandwidth results in a very spiky density estimate.
The lowest value of $\sigma$ that results in correct sampling, $\sigma_{\mathrm{crit}}$, depends on the system and the number of particles. This is investigated further in Section~S-IV in the SM~\cite{supplemental_material}.

We further examine the performance of the different approximations by evaluating how quickly the birth-death algorithm manages to distribute the particles between the two states in the correct ratio according to the equilibrium distribution.
In Fig.~\ref{fig:correction}(d), we show the fraction of particles in left state, $N_{\mathrm{left}}/N$, where $N_{\mathrm{left}}$ is the number of particles in the left state as defined by the basin of attraction $B_L=(-\infty,x_0)$. Note that we only show the initial 8000 steps of the simulations (i.e., the first $0.4\%$ of the total simulation). Additionally, we show results obtained without birth-death events, that is, a pure overdamped Langevin dynamics simulation with the same parameters and number of particles but with the particles moving totally independent, so that the same amount of statistics are used to estimate the energy landscapes.
The reference equilibrium value is $\pi(B_L) \approx 0.63$, in other words, there should be around 63 particles in the left state in the current case of $N=100$. As mentioned above, we start with a particle distribution far from equilibrium as initially there are only 10 particles in the left state.
Without the birth-death process, the simulation only slowly tends towards the equilibrium value and has not reached it within the time frame shown in Fig.~\ref{fig:correction}(d). However, at longer times, the pure Langevin dynamics simulation reaches the correct equilibrium distribution.
This is due to the fact that the moderate barrier height of the system allows for transitions from the Langevin dynamics alone within the simulation time, although the respective time scale of transition is long.
In contrast, all simulations that employ the birth-death scheme quickly approach the correct equilibrium particle distribution and reach the reference equilibrium value $\pi(B_L)\approx 0.63$ with a few thousand Langevin steps.

We can see in Fig.~\ref{fig:correction}(d) that similar results are obtain with our new birth-death term $\Lambda^{\mathrm{mu}}$ and the original birth-death term $\Lambda^{0}$. Therefore, the incorrectness of the original birth-death term $\Lambda^{0}$ is mainly exhibited in the sampling of the barrier region while equilibrium properties seem to be less affected. Furthermore, we can see in Fig.~S5 in the SM~\cite{supplemental_material}, that if we view the results in terms of the probability distribution, the issue with the undersampling of the barrier region with the original birth-death term $\Lambda^{0}$ is barely noticeable. This can explain why this issue was not noticed in Ref.~\citenum{lu_accelerating_2019}, as there the authors only viewed the results in terms of probability distributions. In this context it should be mentioned that obtaining accurate estimates of barriers is an important problem in computational physics and chemistry~\cite{10.1063/5.0020240}, so it is important that the birth-death scheme correctly samples the barrier region.

As can be seen in Fig.~S6 in the SM~\cite{supplemental_material}, we obtain overall similarly good results with the additive birth-death term $\Lambda^{\mathrm{ad}}$ in Eq.~\ref{eq:correction_add} as with the new multiplicative birth-death term $\Lambda^{\mathrm{mu}}$.

As noted in Lemma 1, increasing the kernel bandwidth will gradually decreases the effect of the birth-death process.
However, for the current case, we only observed a significant slow-down of the equilibration for very large kernel widths such as $\sigma = 5$, as shown in Section S-VI in the SM~\cite{supplemental_material}.
In practice, we should therefore focus on choosing $\sigma$ large enough to get a smooth density estimate and correct sampling while keeping in mind that too large bandwidth values reduce the effectiveness of the method.

\subsection{Influence of birth-death stride $M$}\label{sec:appl_bd_stride}
We next examine the effect of the birth-death stride $M$ that determines how many Langevin dynamics steps are performed between birth-death attempts.
To this end, we perform a set of simulations with the same double-well energy landscape as before and vary $M$ while keeping the number of particles and the kernel width fixed at $N=100$ and $\sigma = 0.4$ respectively. All other parameters are the same as in the previous section.
As before, we start with an particle distribution far from equilibrium with 10 percent of the particles in the more probable left state and the rest in the right state.
We use the same protocol as before and obtain estimates of the energy landscape using histogramming.

\begin{figure*}
    \includegraphics{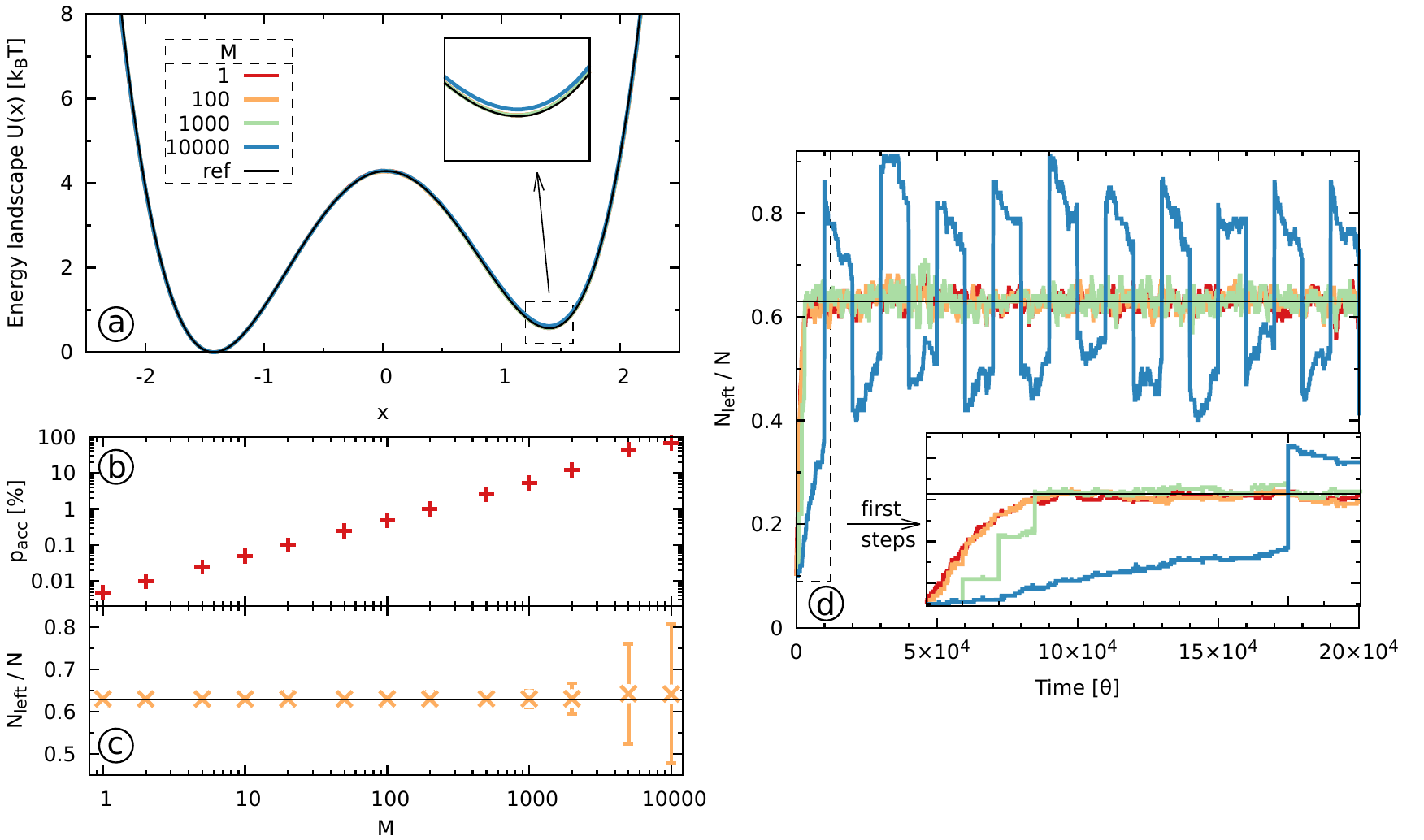}
    \caption{
        (a) Estimates of the energy landscape for the potential given in Eq.~\eqref{eq:double_well} obtained from sampling using histogramming for different values $M$ of Langevin dynamics steps between attempted birth-death events. All but the data for $M=10{,}000$ cannot be distinguished from the reference.
        (b) Percentage of accepted birth-death events of the total number of birth-death attempts.
        (c) Average fraction of particles in the left state. The error bars denote the standard deviation and the black horizontal line is the expected equilibrium value. The first 50,000  steps were omitted when calculating the values.
        (d) Fraction of particles in the left state as a function of simulation time. The colored lines are from the same simulations as in (a), the black horizontal line is the expected equilibrium value. The inset shows a magnification of the first 12,000 steps.
    \label{fig:bd_stride}}
\end{figure*}

The results of the simulations are shown in Fig.~\ref{fig:bd_stride}.
In panel (a), we present the estimated energy landscapes.
In panel (b), we show the percentage of accepted birth-death events $p_{\mathrm{acc}}$ given by the number of executed birth-death moves divided by the total number of attempted ones.
This can be understood as an estimate of the average birth-death probability of Eq.~\eqref{eq:bd_probs} during the simulations.
In panel (d), we show the time evolution of the particle distribution for the two states obtained in the same way as in the previous section by considering the fraction of particles in the left state, $N_{\mathrm{left}}/N$.
Additionally, in panel (d),
we show the mean and standard deviation of the fraction of particles in the left state, calculated by omitting the first 50,000 steps of the equilibration phase.

We can see in panel (a) that all simulations yield a good estimate of the energy landscape, although the one with $M = 10{,}000$ shows a slight deviation at the minimum of the right state as can be seen in the inset.
Looking at the time evolution of the particle distribution in panel (c), we see that for all simulations with $M < 10{,}000$ the correct equilibrium distribution is reached within the first 4,000 steps and there are only small fluctuations around the reference value afterwards.
Smaller $M$ values result in slightly faster equilibration, although we find this effect to be rather small.

For the simulation with $M = 10{,}000$, the birth-death events result in ``overshooting'', such that the number of particles in the left state becomes either too small or too large directly afterward.
Between the birth-death events, a slow equilibration process due to the Langevin dynamics can be observed, as the moderate barrier height makes transitions only rare but not completely unlikely.
The overshooting happens because we calculate the birth-death probabilities for all particles at once and then perform the respective events simultaneously.
The time between birth-death calculations enters exponentially in the event probabilities in Eq.~\ref{eq:bd_probs}.
For large values of $M$, the event probabilities thus become very large, and around $70\,\%$ of the particles are killed or duplicated each time. While a per-particle approach with recalculation of the probabilities after each accepted event would solve the problem, this would also result in a lot more computational effort.
An equivalent simulation with the recalculation of the birth-death probabilities after each birth-death event is shown in Section S-VII in the SM~\cite{supplemental_material}, where we can see that this solves the problem.

The reason for not calculating the probabilities at every Langevin step is to lower the computational effort. Therefore, we conclude that, as long as the birth-death events remain relatively rare, performing multiple Langevin steps between birth-death attempts helps to speed up simulations without negative side effects.
To quantify this for the given system, we observe significant changes in the behavior only for $M>1000$ in Fig.~\ref{fig:bd_stride}(c) which corresponds to $p_{\mathrm{acc}} > 5\,\% $.

\subsection{Underdamped Langevin dynamics, and the effect of barrier height on the speed of equilibration}
\label{sec:underdampedlangevinsimulations}
After investigating the influence of the parameters of the birth-death algorithm, we evaluate the behavior for the underdamped Langevin case that was introduced in Section~\ref{sec:theory_underdamped_langevin}.
To simultaneously assess the speed of convergence for different barrier heights, we generalize
the double-well energy landscape given in Eq.~\eqref{eq:double_well} with two parameters $a$ and $b$:
\begin{equation}\label{eq:double_well_scaled}
  U(x) = ax^4 - 4ax^2 + bx + C,
\end{equation}
where, as before, $C$ is a constant such that $\min U(x) = 0$.
In Table~\ref{tab:double_well_coeffs}, we give sets of coefficients where we have systematically increased $a$ and then set $b$ such that the equilibrium distribution of the particles in the two states remains fixed at about $\pi(B_L) \approx 0.63$ to $\pi(B_R) \approx 0.37$ as it was in the previous sections, while the barrier height is increased.

\begin{table}
    \caption{Coefficients for potentials according to Eq.~\eqref{eq:double_well_scaled} with increasing barrier height while keeping the equilibrium distribution of particles between the two states fixed so that $\pi(B_L) \approx 0.63$.\label{tab:double_well_coeffs}}
    \begin{ruledtabular}
    \begin{tabular}{ cdd }
        $a$ [$k_\mathrm{B}T$] & \multicolumn{1}{c}{$b$ [$k_\mathrm{B}T$]} & \multicolumn{1}{c}{barrier height} [$k_\mathrm{B}T$] \\
        \hline
        $1$ & 0.2 & 4.285 \\
        $2$ & 0.1918 & 8.272 \\
        $4$ & 0.1889 & 16.267 \\
        $8$ & 0.1877 & 32.262\\
    \end{tabular}
    \end{ruledtabular}
\end{table}

For each set of coefficients, we perform simulations with the Langevin solver from Ref.~\citenum{bussi_accurate_2007}.
We set $\beta=1$, $m=1$ and $\gamma=10$ and simulate $N=100$ particles for 2,000,000 steps with a time step of $\theta=0.005$, where we again start with only 10 particles in the left state and 90 in the right state.
We perform birth-death events performed every $M=100$ Langevin steps. We use the birth-death term $\Lambda^{\mathrm{mu}}$ and kernel width $\sigma=0.5$.
For comparison, we additionally perform pure Langevin dynamics simulations without birth-death events but otherwise the same parameters and numbers of particles.
We perform analog analysis of the simulations as in the previous sections by estimating the energy landscapes using histogramming.

\begin{figure*}
    \includegraphics{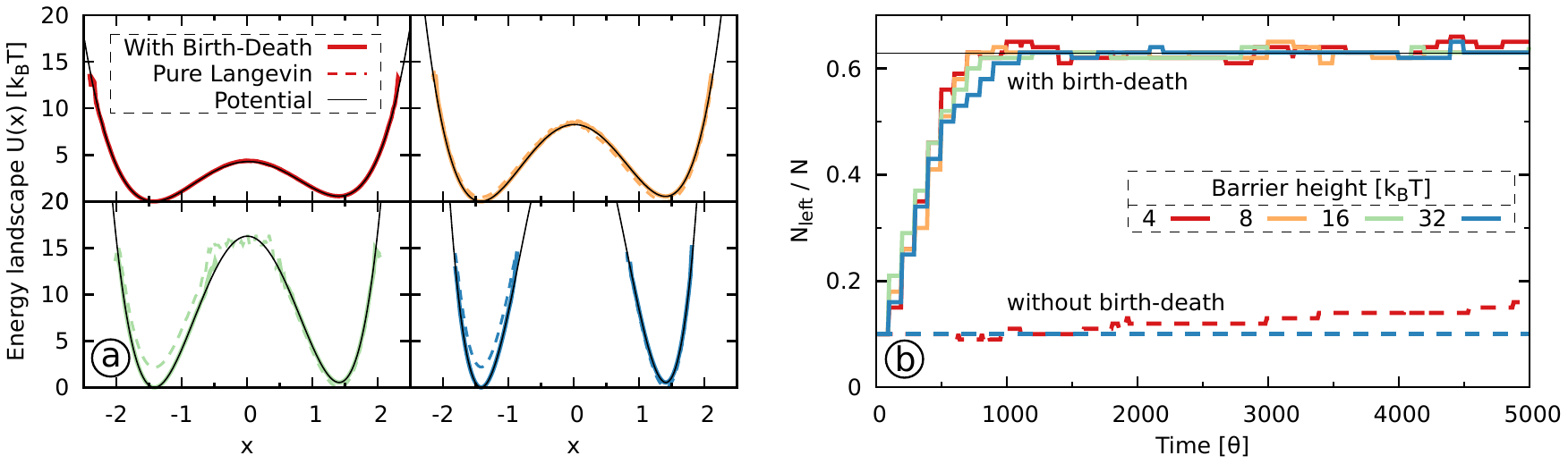}
    \caption{
        (a) Estimates of the energy landscape obtained from sampling using histogramming for the potential given in Eq.~\eqref{eq:double_well_scaled} with different coefficients given in Table~\ref{tab:double_well_coeffs}. The solid lines are from simulations with birth-death events, the dashed lines from Langevin dynamics simulations without birth-death events (i.e., independent particles), and the thin black lines are the reference from the potential.
        (b) Number of particles in the left state as a function of simulation time for the different potentials. Shown are only the first 5,000 steps. The solid black line is the equilibrium value calculated from the potentials (same for all potentials).}
    \label{fig:underdamped_langevin}
\end{figure*}

We show the results of the simulations in Fig.~\ref{fig:underdamped_langevin}.
For the simulations with birth-death events, we can see that the estimated energy landscape agrees well with the reference one in all cases.
Note that for the higher barrier heights, the barrier region is not sampled due to low probability and insufficient simulation time, as can be expected.

On the contrary, we observe that the pure Langevin dynamics simulations without birth-death events are only sampling the system correctly if the barrier heights is low.
Already for the system with a barrier height of $8\:k_{\mathrm{B}} T$, there is a visible difference for the right state that is estimated to be lower in energy than the reference. For higher barriers, the estimates are completely off.
They give the right state as lower in energy than the left one by several $k_{\mathrm{B}} T$ in comparison to the reference energy landscape.

We can see the reason for these results for the pure Langevin dynamics by looking at the distribution of the particles in the two states shown in Fig.~\ref{fig:underdamped_langevin}(b).
With increased barrier height, crossings between the states by pure Langevin dynamics become rarer.
While the probability of crossing the barrier is not zero for the higher barrier heights, transitions are too rare to equilibrate the particles across the two states within the simulation time.
In fact, we could not observe a single transition in the pure  Langevin dynamics simulation for the system with a barrier height of $32\: k_{\mathrm{B}} T$.
Therefore, a pure Langevin dynamics simulation is unable to sample the energy landscape correctly, but this can be expected due to high barrier heights.

On the other hand, the simulations with birth-death events reach the equilibrium distribution of the particles very quickly, that is, within the first 1,000 steps of the simulation.
It is clearly visible that the barrier height has only a negligible influence on the speed of equilibration, which is in accordance with a similar theoretic result for an overdamped system with no smoothing kernel (see Theorem 3.3 of Ref.~\citenum{lu_accelerating_2019}).
Here, this theoretical derivation is found to be also true when using an approximation (in this case $\Lambda^{\mathrm{mu}}$) to the birth-death term.

\subsection{Higher dimensions -- The two-dimensional Wolfe-Quapp potential}
\begin{figure*}
    \includegraphics{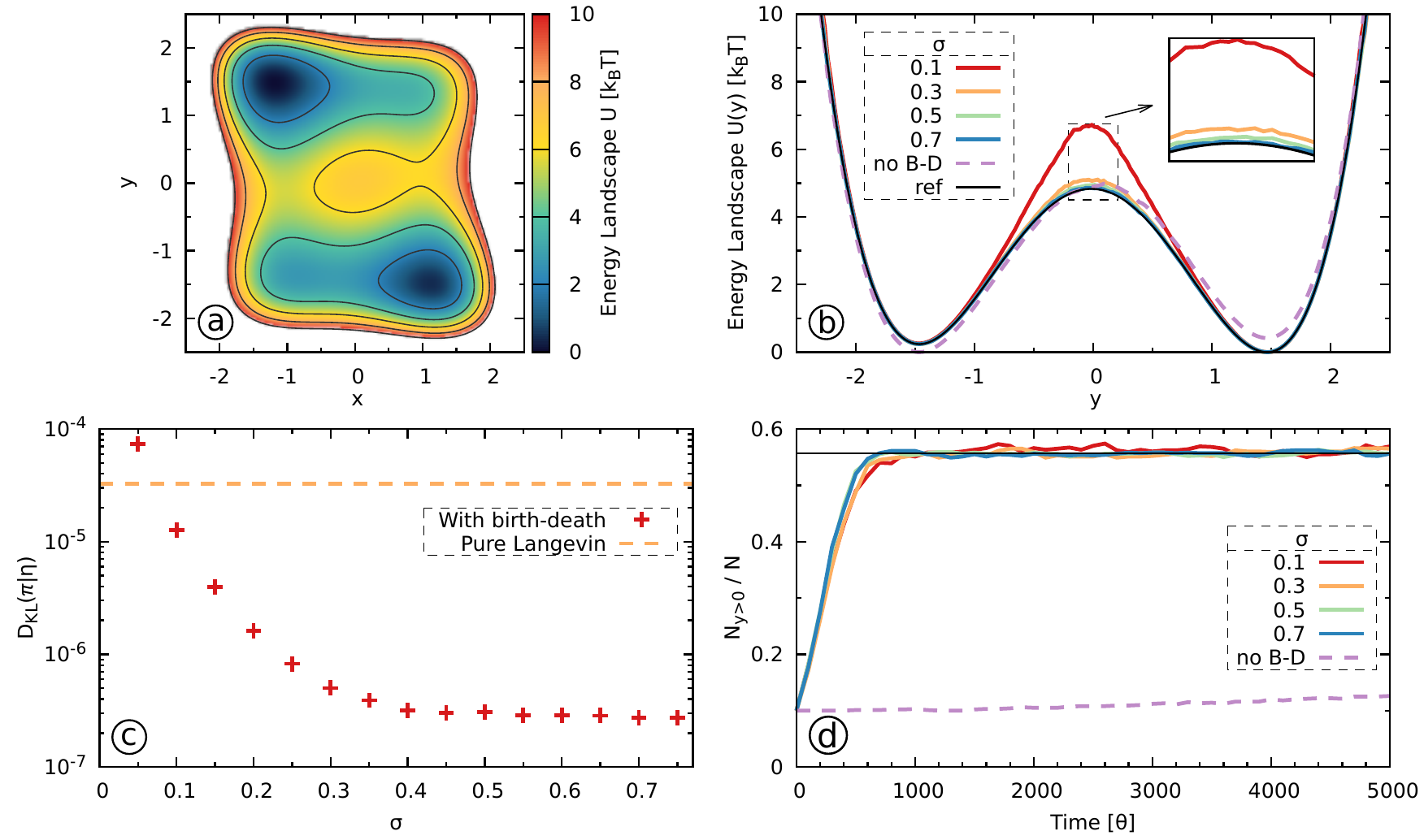}
    \caption{
        (a) The reference energy landscape of the Wolfe-Quapp potential (Eq.~\eqref{eq:wq}).
        (b) The energy landscapes estimated from sampling using histogramming projected on the y-direction. Colored solid lines are from simulations with birth-death events using different kernel widths $\sigma$. We note that the lines for $\sigma=0.55$ and $\sigma=0.75$ are hardly distinguishable because they basically are on top of each other. The dashed line is from a Langevin simulation without birth-death events but the same number of particles. For clarity, the dashed line is omitted in the inset. The black line is the reference energy landscape calculated from the potential.
        (c) Kullback-Leibler divergences from the estimated probability distribution to the equilibrium distribution for simulations with different kernel widths $\sigma$. For comparison, the dashed horizontal line is from a Langevin simulation without the birth-death algorithm.
        (d) Fraction of particles in the state with $y>0$ as a function of simulation time. The black horizontal line is the reference equilibrium value. The different lines represent the same simulations as in (b).
    \label{fig:wq}}
\end{figure*}

All of the previously presented simulations were performed for a system with only one spatial dimension.
As the theory from Section~\ref{sec:theory} holds for higher dimensions, we also test the performance on the two-dimensional Wolfe-Quapp potential~\cite{wolfe_chemical_1975,quapp_growing_2005} given by
\begin{equation}\label{eq:wq}
  U(x,y) = x^4 +y^4 - 2 x^2 -4 y^2 + xy + 0.3 x + 0.1 y + C,
\end{equation}
where, as before, $C$ is a constant such that $\min U(x,y) = 0$.
This energy landscape can be seen in Fig.~\ref{fig:wq}(a).
Transitions between the states in $y$-direction are rare events while the mobility in $x$-direction is high, though the two coordinates are highly coupled.

We run simulations using $N=1{,}000$ particles for 200,000 steps with the underdamped Langevin solver, but otherwise use the same parameters as for the one-dimensional systems in Section~\ref{sec:underdampedlangevinsimulations}.
The initial distribution is again chosen to be far from equilibrium: we place 100 particles in the metastable state in the top left corner with a minimum at (-1.17, 1.48) and 900 particles in the state in the bottom right corner with a minimum at (1.12, -1.49).
The bandwidths of the Gaussian kernel are chosen to be the same in each direction, $\sigma = \sigma_{x} = \sigma_{y}$, because the low-energy region of the potential has roughly the same size in both dimensions.
We note that this is not a requirement and asymmetric kernels can be employed just as well.
The kernel bandwidths are varied in the range $\sigma \in [0.05, 0.75]$ in steps of $0.05$.
Additionally, for comparison, we perform a pure Langevin dynamics simulation without the birth-death events but otherwise the same simulation protocol.

As before, we estimate the energy landscape by histogramming the simulations where we omit the first 10,000 steps.
Additionally, we assess the correctness of the sampling by using the Kullback-Leibler divergence (as defined in Eq.~\eqref{eq:kl_div}) between the equilibrium probability distribution $\pi$ and the estimated distributions $\eta$ obtained from normalizing the histograms $H$ of the simulations. A lower value of the Kullback-Leibler divergence indicates a better agreement of the estimated energy landscapes with the reference ones.

In Fig.~\ref{fig:wq}, we show results of the simulations. As a visual inspection of two-dimensional energy landscapes is difficult, we choose to consider the projection onto the $y$-direction that is shown in panel (b).
We can observe similar results as for the 1D potential: above a certain kernel width $\sigma_{\mathrm{crit}}$, the estimate energy landscapes are very close to the reference. This can also be seen in Kullback-Leibler divergence in panel (c). However, for too narrow kernels, the barrier regions are overestimated and we obtain a higher Kullback-Leibler divergence value.
In panel (d), we can observe, that as before, the birth-death algorithm leads to a very swift equilibration of the  particles between the two metastable states.

We can see that the pure Langevin simulation without birth-death event does not correctly sample the system and significantly deviates not only in the barrier region but also in the relative height of the two basins. This corresponds to the Kullback-Leibler divergence value that is more than one order of magnitude higher than results obtained with the birth-death simulations.

We also performed simulations with a lower number of particles, $N=100$, compared to $N=1000$ in Fig.~\ref{fig:wq}, see results in Section S-VIII.1 in the SM~\cite{supplemental_material}. We obtain similarly good results as for the case with more particles presented here, though the results are slightly more noisy.

Furthermore, we performed simulations with a scaled version of the Wolfe-Quapp potential, such that the barrier heights are increased, see results in Section S-VIII.2 in the SM~\cite{supplemental_material}. Again, we observe there that the birth-death algorithm is able to obtain a correct sampling of the energy landscape, while pure Langevin sampling is unable to obtain good results. We can also see that the speed of equilibration to the correct particle distribution is independent of the barrier height, as observed in the previous section.

\section{Summary and Outlook}
\label{sec:summary}
In this paper, we have explored the usage of the birth-death scheme from Ref.~\citenum{lu_accelerating_2019} to sample rare event energy landscapes. We amend a deficiency of the original algorithm by introducing a new approximate birth-death term that has the right mathematical limits and empirically leads to correct sampling of barrier regions between metastable states. We show empirically that the birth-death scheme can very efficiently sample prototypical rare event energy landscapes, both for overdamped and underdamped Langevin dynamics, and that the speed of equilibration is independent of the barrier height. We also show that the computational effort can be reduced by applying birth-death steps less frequently without negatively affecting the quality of the sampling. This is an important point for future applications where we would like to reduce the communication between the different simulations.

Overall, our results show that this birth-death scheme is a promising sampling method that could extend not only Langevin dynamics but also other sampling schemes.
We provide empirical evidence that the algorithm can also be used with more general sampling schemes, which motivates testing its applicability further, for example to molecular dynamics or Monte Carlo simulations.
However, to be able to apply it to simulations of high-dimensional systems, such as physical and chemical systems, will require considerable future work.
To obtain smooth estimates of the particle distribution without increasing the kernel width (and therefore making the algorithm less efficient), the number of particles has to be increased with the number of dimensions, which makes the algorithm only applicable to a few degrees of freedom.
A future extension could therefore modify the algorithm to perform the birth-death moves only in some relevant subspace, that is, in a low-dimensional space of a few collective variables.
In other words, the birth-death algorithm would treat the particles that represent the simulations as if they were moving on the free energy landscape corresponding to these collective variables instead of the high-dimensional space.
In this case, the free energy landscape, and thus the corresponding probability distribution that enters the birth-death term, is unknown a priori. One would need to estimate the probability distribution on the fly during the simulation, which could be done, for example, by combining the method with collective variable-based enhanced sampling methods. We will address this issue in future work.

\begin{acknowledgments}
We thank Oleksandra Kukharenko and Kostas Daoulas (Max Planck Institute for Polymer Research) for carefully reading over the manuscript. We acknowledge support from the Deutsche Forschungsgemeinschaft (DFG, German Research Foundation) - Project number 233630050 - TRR 146 ``Multiscale Simulation Methods for Soft Matter Systems''.
\end{acknowledgments}

\section*{Author contributions}
B.P.\ and S.H.\ contributed equally to this work.
B.P.\ implemented the algorithm and performed the numerical simulations and corresponding analysis of Section~\ref{sec:applications}.
S.H.\ provided the theoretical part of the paper and developed the proofs of Section~\ref{sec:theory}.
L.H.\ and O.V.\ conceptualized the work and supervised the project.
All authors discussed the results and contributed to writing the final manuscript.

\bibliography{Refs_PampelHolbachHartungValsson}

\begin{thebibliography}{57}
\bibitem [{sup()}]{supplemental_material}%
  \BibitemOpen
  \href@noop {} {\bibinfo {title} {See {S}upplemental {M}aterial at [url will
  be inserted by publisher] for further results and discussions, including;
  test of the assumptions of {T}heorem 2; momentum equilibrium in the general
  {L}angevin case; effect of varying the birth-death rate factor
  $\tau_{\alpha}$; influence of the number of particles {$N$} on the critical
  bandwidth $\sigma_{\mathrm{crit}}$; additional figures for the comparison
  between approximations {$\Lambda$}; behavior for large kernel bandwidths;
  recalculation of birth-death probabilities for {$M=10,000$}; additional
  simulations with the {2D} {W}olfe-{Q}uapp potential; influence of the choice
  of {L}angevin time step $\theta$.}}\BibitemShut {Stop}%
\end{thebibliography}
\end{document}